# Machine Learning Techniques to Address Cybersecurity Challenges

Khatoon Mohammed

Abstract.
The increasing prevalence and complexity of cyber attacks necessitate enhanced malware detection methods to uphold computer system security. Traditional signature-based approaches exhibit limitations in addressing the dynamic nature of modern threats. In this context, machine learning (ML) has emerged as a viable solution, offering superior capabilities in detecting malware. ML algorithms excel in processing large datasets and identifying complex patterns beyond human recognition.

This document provides an exhaustive analysis of cutting-edge ML methodologies applied in malware detection, encompassing supervised and unsupervised learning, deep learning, and reinforcement learning. It also addresses the challenges and constraints of ML-based malware detection, including vulnerability to adversarial attacks and the need for substantial amounts of labeled data.

Moreover, the paper explores prospective advancements in ML-based malware detection, highlighting the potential amalgamation of various ML algorithms and the adoption of explainable AI to improve the interpretability of ML-driven detection systems. The insights from this research underscore the effectiveness of ML approaches in enhancing the speed and precision of malware detection, thereby contributing significantly to the fortification of cybersecurity.

Introduction

Machine learning (ML) has emerged as a promising solution to detect and prevent cyber attacks. Cybersecurity threats have become increasingly frequent and sophisticated in recent years, and traditional methods of threat detection are struggling to keep up with the volume and complexity of these threats. However, ML has the potential to significantly improve the speed and accuracy of threat detection, making it a powerful tool in the fight against cybercrime. One of the key advantages of ML in cybersecurity is its ability to analyze large datasets and identify patterns that are difficult for humans to detect. ML algorithms can learn from previous attacks and adapt to new threats, improving their accuracy over time. This adaptability is crucial in the constantly evolving world of cybersecurity, where attackers are constantly developing new tactics to breach networks and steal sensitive data. ML algorithms can also be used to detect anomalies and patterns in network traffic, which can indicate the presence of a cyber attack. For example, an ML algorithm can be trained to recognize patterns of behavior associated with malware, such as unusually high data transfer rates or frequent attempts to connect to suspicious domains. By identifying these patterns, ML algorithms can quickly alert security teams to the presence of a potential threat, allowing them to take action before significant damage is done. Another application of ML in cybersecurity is in the field of fraud detection. ML algorithms can be trained to recognize patterns of behavior associated with fraudulent activity, such as unusual spending patterns or login attempts from unfamiliar devices. By analyzing large datasets of transactions and user behavior, ML algorithms can quickly identify potential instances of fraud and alert security teams to investigate further. However, ML-based threat detection is not without its challenges. One of the biggest challenges is the need for large amounts of labeled data to train ML algorithms effectively. In cybersecurity, labeled data refers to datasets that have been manually annotated to indicate whether they contain malicious or benign behavior. While there are many public datasets available for cybersecurity researchers to use, they often suffer from a lack of diversity or may not accurately reflect real-world threats. Another challenge is the potential for ML algorithms to be fooled by adversarial attacks. Adversarial attacks involve manipulating data in a way that causes an ML algorithm to misclassify it. For example, an attacker could modify a benign file in a way that causes an ML-based malware detector to classify it as malicious. This is a significant concern in cybersecurity, as it means that ML-based threat detection systems may not be reliable in the face of determined attackers. Despite these challenges, ML has the potential to significantly improve the speed and accuracy of threat detection in cybersecurity. As the volume and complexity of cyber threats continue to increase, it is becoming clear that traditional methods of threat detection are no longer sufficient. ML-based approaches offer a promising solution, allowing security teams to quickly identify potential threats and respond before significant damage is done. One of the most significant developments in ML-based cybersecurity in recent years has been the emergence of explainable AI (XAI) techniques. XAI refers to methods that allow ML algorithms to provide explanations for their decision-making processes. This is important in cybersecurity, where it is often necessary to understand why an ML algorithm has classified a particular behavior as malicious or benign. XAI techniques can also be used to identify vulnerabilities in ML-based threat detection systems. By analyzing the explanations provided by an ML algorithm, security teams can identify cases where the algorithm may be susceptible to adversarial attacks. This allows them to take steps to harden the algorithm against such attacks, improving the overall security of the system. In conclusion, ML has the potential to significantly improve the speed and accuracy of threat detection in cybersecurity. By analyzing large datasets and identifying patterns that are difficult for humans to detect, ML algorithms can quickly alert security teams to the presence of potential threats, allowing them to take action before significant damage is done. However, there are also significant challenges associated with ML
.

The efficacy of employing adversarial examples as a means to enhance the robustness of Natural Language Processing (NLP) models against adversarial attacks has recently been revealed in numerous research endeavors (Guu et al., 2018; Iyyer et al., 2018; Alvarez-Melis and Jaakkola, 2017; Jia and Liang, 2017; Ebrahimi et al., 2018; Naik et al., 2018). It is worth mentioning that Alzantot et al. (2018) and Jin et al. (2020) have devised methodologies for generating adversarial texts by substituting words with their synonyms, as determined by their proximity in the word embedding space. This approach leads to the creation of written content that perplexes models and induces erroneous classifications. Zhao et al. (2018) proposed a novel approach to creating adversarial instances in the context of continuous data representation. Their research focused on exploring the semantic domain and resulted in the development of a Generative Adversarial Network (GAN) that is capable of producing coherent and comprehensible adversarial examples. Jia et al. (2019) conducted a study in a relevant context, whereby they explored word replacement combinations that aim to reduce the largest worst-case loss. This investigation employed the widely utilized interval bound propagation technique. Zhu et al. (2020) have adopted a unique approach by focusing on the incorporation of adversarial perturbations into word embeddings, rather than generating textual outputs directly. The objective of this approach is to mitigate adversarial risk associated with input instances.

Our research builds upon the work of Wang et al. (2020) in the field of controlled adversarial text generation. Specifically, we want to enhance the performance of their model, CAT.Gen, by generating adversarial texts that are both more engaging and coherent. The scope of their suggested model's implementation is constrained to a singular dataset, specifically the Amazon Review dataset, and a singular machine learning architecture, namely the Recurrent Neural Network (RNN). However, the model demonstrates its capability to generate more authentic and contextually relevant adversarial instances in real-world scenarios. The objective of our study is to expand upon their established methodology by applying it to an alternative dataset (IMDB) and employing a transformer-based neural network. This choice is motivated by the remarkable performance exhibited by transformer models, such as BERT, in several language tasks, including sentiment categorization. The inclusion of a grammatical validation process, which ensures that the produced adversarial examples conform to proper grammar and semantic coherence, is a notable enhancement compared to the CAT-Gen model. Furthermore, a notable computational burden arises when training extensive batches of adversarial samples, which is a limitation of the Wang et al. (2020) study. This problem becomes particularly evident when applying their approach to a substantial dataset such as Yelp Polarity. In light of the considerable processing time we encountered, despite utilizing robust computer resources, our investigation focused on identifying a more efficient approach for generating hostile instances. To accomplish this objective, the inner ascent stages of Projected Gradient Descent (PGD) are employed. PGD is a widely used optimization technique in the field of machine learning due to its ability to efficiently compute parameter gradients without incurring significant additional computational costs when computing gradients of inputs.

The investigation we are conducting is closely connected to the topic of controllable text generation. In contrast, Iyyer et al. (2018) developed a system known as the Syntactically Controlled Paraphrase Network (SCPN) to generate adversarial instances. For example, Hu et al. (2017) conducted a study exploring the utilization of variational auto-encoders in combination with holistic attribute discriminators. The approach utilized in this study is founded upon an encoder-decoder model, which effectively generates adversarial training data, hence enhancing the model's resilience against adversarial attacks. Furthermore, Zhu et al. (2019) propose FreeLB, a unique technique for adversarial training that aims to enhance the level of invariance within the embedding space. In this approach, the inclusion of adversarial perturbations in word embeddings is employed, and the subsequent minimization of adversarial risk is carried out across many regions including the input samples. The utilization of this approach on Transformer-based models for tasks pertaining to the comprehension of natural language and reasoning based on common sense demonstrates its effectiveness, as evidenced by the attainment of improved benchmark test outcomes for models such as BERT-base and RoBERTa-large. The authors' treatment of the temporal features of NLP models is insufficient, as they neglect to include potential temporal variations and shifts in data patterns. This issue holds significant importance, particularly in practical scenarios where the forthcoming data may exhibit variations from the characteristics and patterns observed in the training data.

## 3 METHOD

Our proposed strategy for producing adversarial cases is schematically depicted in Figure 1. Our approach is optimized for creating attacks with a narrow focus, in this case sentiment classification. This is accomplished by manipulating the property in question inside a given input sentence (in this case, product reviews). Encoder, decoder, and attribute classifier make up the model architecture; these components are also present in prior work on controlled text generation (Hu et al., 2017; Shen et al., 2017; Dathathri et al., 2020). The system incorporates parts to make it easier to alter attributes and create assaults inside a given job model.

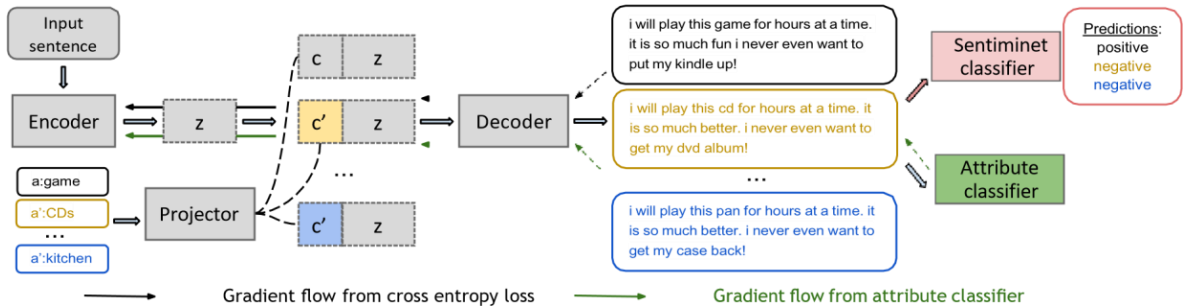

Figure 1 provides an overview of our adversarial examples generation technique. The process of backpropagation is employed. The utilization of cross entropy loss (shown by a black dashed line) is employed to guarantee that the adversarial examples created by our system comply with grammar requirements and maintain semantic coherence. Please guarantee that the sentence generated possesses a comparable semantic meaning to the original input sentence. The manipulation of attributes (shown by the green dashed line) is used to induce attribute loss in the generated sentence, which is unrelated to the task label. The prediction of sentiment labels on generated text exhibits variability while altering the attribute denoted as "a" representing different categories.

| Method | Examples |
|---|---|
| Textfooler (Jin et al., 2020) | A person is relaxing on his day off → A person is relaxing on his nowadays off<br>The two men are friends → The three men are dudes |
| NL-adv (Alzantot et al., 2018) | A man is talking to his wife over his phone → A guy is chitchat to his girl over his phone<br>A skier gets some air near a mountain... → A skier gets some airplane near a mountain... |
| Natural-GAN (Zhao et al., 2018) | a girl is playing at a looking man . → a white preforming is lying on a beach .<br>two friends waiting for a family together . → the two workers are married . |

Examples of adversarial text generation models that have been tested on the SNLI dataset are provided in Table 1 above (Bowman et al., 2015). Word-swapping techniques (Textfooler and NL-adv) can generate adversarial text with diverse or restricted semantics, while GAN techniques (Natural-GAN) are more likely to produce sentences that break the rules of the adversarial task.

### 3.1 The generation of adversarial examples at scale with high efficiency.

It is firmly believed that the optimization process is fundamental to any machine learning algorithm. Consequently, significant effort was dedicated to improving our algorithm in order to efficiently generate substantial batches of adversarial examples (AEs) and ultimately develop the most effective adversarial attacks. In order to achieve this objective, The inner ascent stages of Projected Gradient Descent (PGD), which is a well-established and highly efficient optimization technique, were utilized in our study. in the field of machine learning. By utilizing PGD, we were able to extract the gradients of the parameters with minimal computational cost while calculating the gradients of the inputs.

**Algorithm 1** "Free" Large-Batch Adversarial Training (FreeLB-$K$)

**Require:** Training samples $X = \{(Z, y)\}$, perturbation bound $\epsilon$, learning rate $\tau$, ascent steps $K$, ascent step size $\alpha$
1: Initialize $\theta$
2: **for** epoch $= 1 \ldots N_{ep}$ **do**
3:     **for** minibatch $B \subset X$ **do**
4:         $\delta_0 \leftarrow \frac{1}{\sqrt{N_\delta}} U(-\epsilon, \epsilon)$
5:         $g_0 \leftarrow 0$
6:         **for** $t = 1 \ldots K$ **do**
7:             Accumulate gradient of parameters $\theta$
8:             $g_t \leftarrow g_{t-1} + \frac{1}{K} \mathbb{E}_{(Z,y) \in B} [\nabla_\theta L(f_\theta(X + \delta_{t-1}), y)]$
9:             Update the perturbation $\delta$ via gradient ascend
10:            $g_{adv} \leftarrow \nabla_\delta L(f_\theta(X + \delta_{t-1}), y)$
11:            $\delta_t \leftarrow \Pi_{\|\delta\|_F \leq \epsilon} (\delta_{t-1} + \alpha \cdot g_{adv} / \|g_{adv}\|_F)$
12:         **end for**
13:         $\theta \leftarrow \theta - \tau g_K$
14:     **end for**
15: **end for**

Figure 2: Optimization algorithm details to generate adversarial examples at scale

## 4 Analysis and Discussion

To accomplish our goal, we employ the gong2018adversarial dataset sourced from IMDB. This dataset consists of binarized ratings that incorporate both positive and negative feelings. The dataset was partitioned into a training set and a test set, consisting of 25,000 reviews each. However, only 2,000 reviews from the training set were utilized for the purposes of development and testing. To allow parameter change and final evaluation, a development set and a test set are maintained, each including 10,000 occurrences. Subsequently, our classifier is trained and optimized through the application of the gradient descent optimization algorithm, utilizing both the training and development sets. The evaluation of the model's performance includes measuring its accuracy on the original instances from the test sets, as well as on adversarial examples that are generated using targeted attack methods particularly tailored for the test set. The BERT model, which is considered state-of-the-art (SOTA), will be used to classify both the attributes (category) and the task labels (sentiment) in our text. The projector employs a singular layer of Multilayer Perceptron (MLP). Throughout the progression of our research, we have seen that the training process can exhibit instability as a result of the utilization of the gumbel softmax technique for generating soft embeddings. Additionally, we have noticed that there are instances where the output sentence tends to replicate the input sentence. In accordance with the suggestions put forth by Hu et al. (2017), we meticulously calibrated the temperature parameter for the gumbel softmax. Additionally, it was shown that utilizing a network with limited capacity, such as a one-layer MLP with a hidden size of 256, as the projector for the controlled attribute, along with a higher dropout ratio on sentence embeddings, such as 0.5, contributes to the stabilization of the training process. Both of these discoveries are located in the preceding section. Table 4 presents an analysis of the transferability of our instances in relation to established adversarial text generation systems, namely Jin et al. (2020) and Alzantot et al. (2018). The aforementioned experiments were conducted by Jin et al. and Alzantot et al.

Challenges and Limitations:

Despite the promising results achieved by ML-based malware detection, there are several challenges and limitations that need to be addressed.

References

Chadha, R. D. (2015). Vehicular Ad hoc Network (VANETs): A Review. International Journal of Innovative Research in Computer and Communication Engineering, 3(3), 2339-2346. https://www.rroij.com/open-access/vehicular-ad-hoc-network-vanets-areview.pdf

Das, R., & Morris, T. H. (2017). Machine learning and cyber security. 2017 International

Conference on Computer, Electrical & Communication Engineering (ICCECE).

https://doi.org/10.1109/iccece.2017.8526232

Dua, S., & Du, X. (2016). Data mining and machine learning in cybersecurity. CRC Press.

Federal Bureau of Investigation (FBI). (2021). Internet Crime Report 2021.

https://www.ic3.gov/Media/PDF/AnnualReport/2021_IC3Report.pdf

Ford, V., & Siraj, A. (2014). Applications of Machine Learning in Cyber Security. ISCA 27th International Conference on Computer Applications in Industry and Engineering (CAINE-2014).

https://vford.me/papers/Ford%20Siraj%20Machine%20Learning%20in%20Cyber%20Security%20final%20manuscript.pdf

Grant, M. J., & Booth, A. (2009). A typology of reviews: An analysis of 14 review types and associated methodologies. Health Information & Libraries Journal, 26(2), 91-108.

https://doi.org/10.1111/j.1471-1842.2009.00848.x

Halder, S., & Ozdemir, S. (2018). Hands-on machine learning for cybersecurity: Safeguard your system by making your machines intelligent using the Python ecosystem. Packt Publishing.

Handa, A., Sharma, A., & Shukla, S. K. (2019). Machine learning in cybersecurity: A review. WIREs Data Mining and Knowledge Discovery, 9(4).

https://doi.org/10.1002/widm.1306

Johnson, J. (2021, March 18). Cybercrime: Reported damage to the IC3 2020. Statista. Retrieved May 4, 2022, from https://www.statista.com/statistics/267132/totaldamage-caused-by-by-cyber-crime-in-the-us/

Kaspersky. (2021, June 17). What is a zero-day attack? - Definition and explanation. www.kaspersky.com. Retrieved May 4, 2022, from https://www.kaspersky.com/resource-center/definitions/zero-day-exploit

Mathew, A. (2021). Machine learning in cyber-security threats. SSRN Electronic Journal. https://doi.org/10.2139/ssrn.3769194

Musser, M., & Garriott, A. (2021). Machine Learning and Cybersecurity: Hype and Reality. Center for Security and Emerging Technology (CSET).

https://cset.georgetown.edu/wp-content/uploads/Machine-Learning-and-Cybersecurity.pdf

National Academies of Sciences (NAS). (2020). Implications of artificial intelligence for cybersecurity: Proceedings of a workshop. National Academies Press.

Rupp, M. (2022, March 4). Machine learning and cyber security: An introduction. VMRay. Retrieved May 4, 2022, from https://www.vmray.com/cyber-security-blog/machinelearning-and-cyber-security-an-introduction/

Salloum, S. A., Alshurideh, M., Elnagar, A., & Shaalan, K. (2020). Machine learning and deep learning techniques for cybersecurity: A review. Advances in Intelligent Systems and Computing, 50-57. https://doi.org/10.1007/978-3-030-44289-7_5

Sava, J. A. (2022, February 14). Cyber security attack impact on businesses 2021. Statista. Retrieved May 4, 2022, from https://www.statista.com/statistics/1255679/cybersecurity-impact-on-businesses/

Dhurandhar, A., Tandon, N., Jain, R., & Varshney, P. K. (2020). Machine learning for cybersecurity: a review. ACM Computing Surveys, 53(2), 1-36.

Kaspersky. (2021). Kaspersky Security Bulletin 2020: Statistics. Retrieved from https://www.kaspersky.com/blog/security-bulletin-2020-stats/38154/

Khan, M. M., Hussain, M., Chen, H., & Salah, K. (2020). A survey of machine learning in cybersecurity: Advances, challenges, and opportunities. Journal of Network and Computer Applications, 168, 102711.

Kolosnjaji, B., Demontis, A., Biggio, B., Maiorca, D., Arp, D., & Rieck, K. (2019). Adversarial malware binaries: Evading deep learning for malware detection in executables. In Proceedings of the 34th Annual Computer Security Applications Conference (pp. 84-97).

NIST. (2020). National Cybersecurity Career Awareness Week: Cybersecurity Jobs in Demand. Retrieved from https://www.nist.gov/blogs/cybersecurity-insights/nationalcybersecurity-career-awareness-week-cybersecurity-jobs-demand

Papernot, N., McDaniel, P., Goodfellow, I., Jha, S., Celik, Z. B., & Swami, A. (2018). Practical black-box attacks against machine learning. In Proceedings of the 2018

ACM on Asia Conference on Computer and Communications Security (pp. 506-519).

Scott Alfeld, Xiaojin Zhu, and Paul Barford. Data poisoning attacks against autoregressive models. In Proceedings of the AAAI Conference on Artificial Intelligence, volume 30, 2016.

Adel R Alharbi, Mohammad Hijji, and Amer Aljaedi. Enhancing topic clustering for arabic security news based on k-means and topic modelling.

IET Networks, 10(6):278–294, 2021.

Nabiha Asghar. Yelp dataset challenge: Review rating prediction. arXiv preprint arXiv:1605.05362, 2016.

Nicholas Carlini and David Wagner. Towards evaluating the robustness of neural networks. In 2017 ieee symposium on security and privacy (sp), pages 39–57. IEEE, 2017.

Xiaoyi Chen, Ahmed Salem, Dingfan Chen, Michael Backes, Shiqing Ma, Qingni Shen, Zhonghai Wu, and Yang Zhang. Badnl: Backdoor attacks against nlp models with semantic-preserving improvements. In Annual Computer Security Applications Conference, pages 554–569, 2021.

Xinyun Chen, Chang Liu, Bo Li, Kimberly Lu, and Dawn Song. Targeted backdoor attacks on deep learning systems using data poisoning. arXiv preprint arXiv:1712.05526, 2017.

Leilei Gan, Jiwei Li, Tianwei Zhang, Xiaoya Li, Yuxian Meng, Fei Wu, Shangwei Guo, and Chun Fan. Triggerless backdoor attack for nlp tasks with clean labels. arXiv preprint arXiv:2111.07970, 2021.

Alex Graves. Long short-term memory. Supervised sequence labelling with recurrent neural networks, pages 37–45, 2012.

Guanyang Liu, Mason Boyd, Mengxi Yu, S Zohra Halim, and Noor Quddus.

Identifying causality and contributory factors of pipeline incidents by employing natural language processing and text mining techniques. Process Safety and Environmental Protection, 152:37–46, 2021.

Yingqi Liu, Shiqing Ma, Yousra Aafer, Wen-Chuan Lee, Juan Zhai, Weihang Wang, and Xiangyu Zhang. Trojaning attack on neural networks. 2017.

Yuntao Liu, Yang Xie, and Ankur Srivastava. Neural trojans. In 2017 IEEE International Conference on Computer Design (ICCD), pages 45–48. IEEE, 2017.

Xiaohan Ma, Rize Jin, Joon-Young Paik, and Tae-Sun Chung. Large scale text classification with efficient word embedding. In International Conference on Mobile and Wireless Technology, pages 465–469. Springer, 2017.

Seyed-Mohsen Moosavi-Dezfooli, Alhussein Fawzi, Omar Fawzi, and Pascal Frossard. Universal adversarial perturbations. In Proceedings of the IEEE conference on computer vision and pattern recognition, pages 1765–1773, 2017.

Daowan Peng, Zizhong Chen, Jingcheng Fu, Shuyin Xia, and Qing Wen. Fast kmeans clustering based on the neighbor information. In 2021 International Symposium on Electrical, Electronics and Information Engineering, pages 551–555, 2021.

Clifton Poth, Jonas Pfeiffer, Andreas R"uckl'e, and Iryna Gurevych. What to pretrain on? Efficient intermediate task selection. In Proceedings of the 2021 Conference on Empirical Methods in Natural Language Processing, pages 10585– 10605, Online and Punta Cana, Dominican Republic, November 2021. Association for Computational Linguistics.

Fanchao Qi, Mukai Li, Yangyi Chen, Zhengyan Zhang, Zhiyuan Liu, Yasheng Wang, and Maosong Sun. Hidden killer: Invisible textual backdoor attacks with syntactic trigger. arXiv preprint arXiv:2105.12400, 2021.

Kun Shao, Yu Zhang, Junan Yang, Xiaoshuai Li, and Hui Liu. The triggers that open the nlp model backdoors are hidden in the adversarial samples. Computers & Security, page 102730, 2022.

Jacob Steinhardt, Pang Wei W Koh, and Percy S Liang. Certified defenses for data poisoning attacks. Advances in neural information processing systems, 30, 2017.

Lichao Sun. Natural backdoor attack on text data. arXiv preprint arXiv:2006.16176, 2020.

M Onat Topal, Anil Bas, and Imke van Heerden. Exploring transformers in natural language generation: Gpt, bert, and xlnet. arXiv preprint arXiv:2102.08036, 2021.

Florian Tramèr, Alexey Kurakin, Nicolas Papernot, Ian Goodfellow, Dan Boneh, and Patrick McDaniel. Ensemble adversarial training: Attacks and defenses. arXiv preprint arXiv:1705.07204, 2017.

Huang Xiao, Battista Biggio, Gavin Brown, Giorgio Fumera, Claudia Eckert, and Fabio Roli. Is feature selection secure against training data poisoning? In international conference on machine learning, pages 1689–1698. PMLR, 2015.

Xiang Zhang, Junbo Jake Zhao, and Yann LeCun. Character-level convolutional networks for text

classification. In NIPS, 2015.

YanPing Zhao and XiaoLai Zhou. K-means clustering algorithm and its improvement research. In Journal of Physics: Conference Series, volume 1873, page 012074. IOP Publishing, 2021.

Wright, J., Dawson, M. E., Jr, & Omar, M. (2012). Cyber security and mobile threats: The need for antivirus applications for smart phones. Journal of Information Systems Technology and Planning, 5(14), 40–60.

Omar , M., & Dawson, M. (2013). Research in progress-defending android smartphones from malware attacks. 2013 Third International Conference on Advanced Computing and Communication Technologies (ACCT), 288–292. IEEE.

Dawson, Maurice, Al Saeed, I., Wright, J., & Omar, M. (2013). Technology enhanced learning with open source software for scientists and engineers. INTED2013 Proceedings, 5583–5589. IATED.

Omar, M. (2012). Smartphone Security: Defending Android-based Smartphone Against Emerging Malware Attacks. Colorado Technical University.

Dawson, Maurice, Omar, M., & Abramson, J. (2015). Understanding the methods behind cyber terrorism. In Encyclopedia of Information Science and Technology, Third Edition (pp. 1539–1549). IGI Global.

Fawzi, D. R. A. J., & Omar, M. (n.d.). NEW INSIGHTS TO DATABASE SECURITY AN EFFECTIVE AND INTEGRATED APPROACH TO APPLYING ACCESS CONTROL MECHANISMS AND CRYPTOGRAPHIC CONCEPTS IN MICROSOFT ACCESS ENVIRONMENTS.

Dawson, Maurice, Omar, M., Abramson, J., & Bessette, D. (2014). The future of national and international security on the internet. In Information security in diverse computing environments (pp. 149–178). IGI Global.

Omar, M. (2015b). Insider threats: Detecting and controlling malicious insiders. In New Threats and Countermeasures in Digital Crime and Cyber Terrorism (pp. 162–172). IGI Global.

Dawson, Maurice, Wright, J., & Omar, M. (2015). Mobile devices: The case for cyber security hardened systems. In New Threats and Countermeasures in Digital Crime and Cyber Terrorism (pp. 8–29). IGI Global.

Dawson, Maurice. (2015). New threats and countermeasures in digital crime and cyber terrorism. IGI Global.

Hamza, Y. A., & Omar, M. D. (2013). Cloud computing security: abuse and nefarious use of cloud computing. Int. J. Comput. Eng. Res, 3(6), 22–27.

Omar, M. (2015a). Cloud Computing Security: Abuse and Nefarious Use of Cloud Computing. In Handbook of Research on Security Considerations in Cloud Computing (pp. 30–38). IGI Global.

Davis, L., Dawson, M., & Omar, M. (2016). Systems Engineering Concepts with Aid of Virtual Worlds and Open Source Software: Using Technology to Develop Learning Objects and Simulation Environments. In Handbook of Research on 3-D Virtual Environments and Hypermedia for Ubiquitous Learning (pp. 483–509). IGI Global.

Mohammed, D., Omar, M., & Nguyen, V. (2017). Enhancing Cyber Security for Financial Industry through Compliance and Regulatory Standards. In Security Solutions for Hyperconnectivity and the Internet of Things (pp. 113–129). IGI Global.

Dawson, Maurice, Omar, M., Abramson, J., Leonard, B., & Bessette, D. (2017). Battlefield Cyberspace: Exploitation of Hyperconnectivity and Internet of Things. In Developing NextGeneration Countermeasures for Homeland Security Threat Prevention (pp. 204–235). IGI Global.

Nguyen, V., Omar, M., & Mohammed, D. (2017). A Security Framework for Enhancing User Experience. International Journal of Hyperconnectivity and the Internet of Things (IJHIoT), 1(1), 19–28.

Omar, M., Mohammed, D., & Nguyen, V. (2017). Defending against malicious insiders: a conceptual framework for predicting, detecting, and deterring malicious insiders. International Journal of Business Process Integration and Management, 8(2), 114–119.

Dawson, Maurice, Eltayeb, M., & Omar, M. (2016). Security solutions for hyperconnectivity and the Internet of things. IGI Global.

Omar, M. (n.d.-b). Latina Davis Morgan State University 1700 E Cold Spring Ln. Baltimore, MD 21251, USA E-mail: latinaedavis@ hotmail. com.

Dawson, Maurice, Davis, L., & Omar, M. (2019). Developing learning objects for engineering and science fields: using technology to test system usability and interface design. International Journal of Smart Technology and Learning, 1(2), 140–161.

Banisakher, M., Mohammed, D., & Omar, M. (2018). A Cloud-Based Computing Architecture Model of Post-Disaster Management System. International Journal of Simulation--Systems, Science & Technology, 19(5).

Nguyen, V., Mohammed, D., Omar, M., & Banisakher, M. (2018). The Effects of the FCC Net Neutrality Repeal on Security and Privacy. International Journal of Hyperconnectivity and the Internet of Things (IJHIoT), 2(2), 21–29.

Omar, M., Mohammed, D., Nguyen, V., Dawson, M., & Banisakher, M. (2021). Android application security. In Research Anthology on Securing Mobile Technologies and Applications (pp. 610–625). IGI Global.

Banisakher, M., Omar, M., & Clare, W. (2019). Critical Infrastructure-Perspectives on the Role of Government in Cybersecurity. Journal of Computer Sciences and Applications, 7(1), 37–42.

Omar, M., & Others. (2019). A world of cyber attacks (a survey).

Mohammed, D., Omar, M., & Nguyen, V. (2018). Wireless sensor network security: approaches to detecting and avoiding wormhole attacks. Journal of Research in Business, Economics and Management, 10(2), 1860–1864.

Banisakher, M., Omar, M., Hong, S., & Adams, J. (2020). A human centric approach to data fusion in post-disaster management. J Business Manage Sci, 8(1), 12–20.

Nguyen, V., Mohammed, D., Omar, M., & Dean, P. (2020). Net neutrality around the globe: A survey. 2020 3rd International Conference on Information and Computer Technologies (ICICT), 480–488. IEEE.

Dawson, M., Omar, M., Abramson, J., & Bessette, D. (2014). INFORMATION SECURITY IN DIVERSE COMPUTING ENVIRONMENTS.

Zangana, H. M., & Omar, M. (2020). Threats, Attacks, and Mitigations of Smartphone Security. Academic Journal of Nawroz University, 9(4), 324–332.

Omar, M. (2021c). New insights into database security: An effective and integrated approach for applying access control mechanisms and cryptographic concepts in Microsoft Access environments.

Omar, M. (2021b). Developing Cybersecurity Education Capabilities at Iraqi Universities.

Omar, M. (2021a). Battlefield malware and the fight against cyber crime.

Omar, M., Gouveia, L. B., Al-Karaki, J., & Mohammed, D. (2022). ReverseEngineering Malware. In Cybersecurity Capabilities in Developing Nations and Its Impact on Global Security (pp. 194–217). IGI Global.

Omar, M. (2022e). Machine Learning for Cybersecurity: Innovative Deep Learning Solutions. Springer Brief. https://link.springer.com/book/978303115.

Omar, M. (n.d.-a). Defending Cyber Systems through Reverse Engineering of Criminal Malware Springer Brief. https://link.springer.com/book/9783031116278.

Omar, M. (2022b). Behavioral Analysis Principles. In Defending Cyber Systems through Reverse Engineering of Criminal Malware (pp. 19–36). Springer International Publishing Cham.

Omar, M. (2022g). Principles of Code-Level Analysis. In Defending Cyber Systems through Reverse Engineering of Criminal Malware (pp. 37–54). Springer International Publishing Cham.

Omar, M. (2022d). Introduction to the Fascinating World of Malware Analysis. In Defending Cyber Systems through Reverse Engineering of Criminal Malware (pp. 1–7). Springer International Publishing Cham.

Omar, M. (2022h). Static Analysis of Malware. In Defending Cyber Systems through Reverse Engineering of Criminal Malware (pp. 9–17). Springer International Publishing Cham.

Omar, M. (2022f). Malware Anomaly Detection Using Local Outlier Factor Technique. In Machine Learning for Cybersecurity: Innovative Deep Learning Solutions (pp. 37–48). Springer International Publishing Cham.

Omar, M. (2022a). Application of Machine Learning (ML) to Address Cybersecurity Threats. In Machine Learning for Cybersecurity: Innovative Deep Learning Solutions (pp. 1–11). Springer International Publishing Cham.

Burrell, D. N., Nobles, C., Cusak, A., Omar, M., & Gillesania, L. (2022). Cybercrime and the Nature of Insider Threat Complexities in Healthcare and Biotechnology Engineering Organizations. Journal of Crime and Criminal Behavior, 2(2), 131–144.

Omar, M. (2022c). Defending Cyber Systems Through Reverse Engineering of Criminal Malware. Springer.

Nobles, C. (2021). Banking Cybersecurity Culture Influences on Phishing Susceptibility. Temple University.

Burrell, D. N., & Nobles, C. (2018). Recommendations to develop and hire more highly qualified women and minorities cybersecurity professionals. International Conference on Cyber Warfare and Security, 75–81. Academic Conferences International Limited.

Aybar, L., Singh, G., Shaffer, A., Bahşi, H., Joseph, C., Udokwu, U. T., … Others. (n.d.). Paper Title Author (s) Page.

Nobles, C. (2018a). Botching human factors in cybersecurity in business organizations. HOLISTICA-Journal of Business and Public Administration, 9(3), 71–88.

Burrell, D. N., Aridi, A. S., McLester, Q., Shufutinsky, A., Nobles, C., Dawson, M., & Muller, S. R. (2021). Exploring System Thinking Leadership Approaches to the Healthcare Cybersecurity Environment. International Journal of Extreme Automation and Connectivity in Healthcare (IJEACH), 3(2), 20–32.

Burrell, D. N., Burton, S. L., Nobles, C., Dawson, M. E., & McDowell, T. (2020). Exploring technological management innovations that include artificial intelligence and other innovations in global food production. International Journal of Society Systems Science, 12(4), 267–285.

Burrell, D. N., Courtney-Dattola, A., Burton, S. L., Nobles, C., Springs, D., & Dawson, M. E. (2020). Improving the quality of "The internet of things" instruction in technology management, cybersecurity, and computer science. International Journal of Information and Communication Technology Education (IJICTE), 16(2), 59–70.

Burrell, D. N., Aridi, A. S., & Nobles, C. (2018). The critical need for formal leadership development programs for cybersecurity and information technology professionals. International Conference on Cyber Warfare and Security, 82–91. Academic Conferences International Limited.

Nobles, C., & Burrell, D. (2018a). The significance of professional associations: Addressing the cybersecurity talent gap.

Nobles, C., & Burrell, D. (2018b). Using Cybersecurity Communities of Practice (CoP) to Support Small and Medium Businesses. ICIE 2018 6th International Conference on Innovation and Entrepreneurship: ICIE, 333.

Nobles, C. (2019c). Establishing human factors programs to mitigate blind spots in cybersecurity.

MWAIS 2019 Proceedings, 22.

Burrell, D. N., Nobles, C., Dawson, M., McDowell, T., Hines, A. M., & Others. (2018). A public policy discussion of food security and emerging food production management technologies that include drones, robots, and new technologies. Perspectives of Innovations, Economics and Business, 18(2), 71–87.

Nobles, C. (2015). Exploring pilots' experiences of integrating technologically advanced aircraft within general aviation: A case study. Northcentral University.

Burrell, D. N., Dattola, A., Dawson, M. E., & Nobles, C. (2022). A practical exploration of cybersecurity faculty development with microteaching. In Research Anthology on Advancements in Cybersecurity Education (pp. 477–490). IGI Global.

Nobles, C. (2016). Mid Morning Concurrent Sessions: Human Factors: Human Error and Cockpit Automation: Presentation: Exploring Pilots' Experiences of Integrating Technologically Advanced Aircraft Within General Aviation.

Nobles, C. (2019a). Cyber threats in civil aviation. In Emergency and Disaster Management: Concepts, Methodologies, Tools, and Applications (pp. 119–141). IGI Global.

Nobles, C. (2018c). The cyber talent gap and cybersecurity professionalizing. International Journal of Hyperconnectivity and the Internet of Things (IJHIoT), 2(1), 42–51.

Nobles, C. (2019b). Disrupting the US National Security Through Financial Cybercrimes. International Journal of Hyperconnectivity and the Internet of Things (IJHIoT), 3(1), 1–21.

Burrell, D. N., Dawson, M. E., & Nobles, C. (2020). Innovative doctorate programs in cybersecurity, engineering, and technology in the USA and UK that can be completed by professionals around the world without relocation. ICRMAT, 1–3.

Nobles, C. (2018b). Shifting the human factors paradigm in cybersecurity. Avail-Able at: Https://Csrc. Nist. Gov/CSRC/Media/Events/Federal-Information-Systems-Security-Edu Cators-As/Documents/17. Pdf.

McLester, Q., Burrell, D. N., Nobles, C., & Castillo, I. (2021). Advancing Knowledge About Sexual Harassment Is a Critical Aspect of Organizational Development for All Employees. International Journal of Knowledge-Based Organizations (IJKBO), 11(4), 48–60.

Nobles, C., Tank, N. A. T., & Galmai, K. (n.d.). The 12th Annual Dupont Summit on Science, Technology, and Environmental Policy.

Nobles, C., Tank, N. A. T., & Williams, J. J. (n.d.). This Type of Epidemic in Healthcare Requires an Information Security Plan.

Nobles, C., Nichols, K. W., Burrell, D. N., & Reaves, A. (2022). The Good, Bad, and Ugly of Remote Work During COVID-19: A Qualitative Study. International Journal of Smart Education and Urban Society (IJSEUS), 13(1), 1–18.

Nobles, C., Burrell, D., & Waller, T. (2022). The Need for a Global Aviation Cybersecurity Defense Policy. Land Forces Academy Review, 27(1), 19–26.

Burrell, D. N., Sabie-Aridi, A. S., Shufutinsky, A., Wright, J. B., Nobles, C., & Dawson, M. (2022). Exploring Holistic Managerial Thinking to Better Manage Healthcare Cybersecurity. International Journal of Health Systems and Translational Medicine (IJHSTM), 2(1), 1–13.

Nobles, C. (2002). Investigating Cloud Computing Misconfiguration Errors using the Human Factors Analysis and Classification System. Scientific Bulletin, 27(1), 59–66.

Nobles, C. (2022). Stress, Burnout, and Security Fatigue in Cybersecurity: A Human Factors Problem. HOLISTICA--Journal of Business and Public Administration, 13(1), 49–72.

Burrell, D. N., Aridi, A. S., Nobles, C., & Richardson, K. (2022). An Action Research Case Study Concerning Deaf and Hard of Hearing Diversity and Inclusion in Healthcare Cybersecurity Consulting Organizations. International Journal of Smart Education and Urban Society (IJSEUS), 13(1), 1–12.

Burrell, D. N., Aridi, A. S., Wright, J. B., Nobles, C., Richardson, K., Lewis, E., & Kemp, R. E. (2022). A Case Study Analysis of Pregnancy Discrimination and Women-Friendly Workplaces in US Engineering and Technical Organizations. International Journal of Applied Management Sciences and Engineering (IJAMSE), 9(1), 1–13.

Burrell, D. N., & Nobles, C. (2022). Discovering the Emergence of Technical Sociology in Human Capital Systems and Technology-Driven Organizations. International Journal of Human Capital and Information Technology Professionals (IJHCITP), 13(1), 1–15.

Sabie-Aridi, A. S., Burrell, D. N., Nobles, C., Richardson, K., & Kemp, R. E. (2022). Challenging and Changing Discrimination Against Asian-American Cybersecurity and Engineering Employees in Information Technology Workplaces. International Journal of Sociotechnology and Knowledge

Development (IJSKD), 14(1), 1–17.

Nobles, C., Robinson, N., Cunningham, M., Robinson, N., Cunningham, M., & Cunningham, M. (2022). Straight From the Human Factors Professionals' Mouth: The Need to Teach Human Factors in Cybersecurity. Proceedings of the 23rd Annual Conference on Information Technology Education, 157–158.

Burrell, D. N., Nobles, C., Dawson, M., Lewis, E. J. M., Muller, S. R., Richardson, K., & Aridi, A. S. (2022). Innovative Legitimate Non-Traditional Doctorate Programs in Cybersecurity, Engineering, and Technology. In Applications of Machine Learning and Deep Learning for Privacy and Cybersecurity (pp. 175–188). IGI Global.

Nobles, C., Burrell, D. N., Waller, T., & Cusak, A. (2022). Food Sustainability, Cyber-Biosecurity, Emerging Technologies, and Cybersecurity Risks in the Agriculture and Food Industries. International Journal of Environmental Sustainability and Green Technologies (IJESGT), 13(1), 1–17.

Richardson, K., Burrell, D. N., Lewis, E. J., Nobles, C., Wright, J. B., Sabie-Aridi, A. S., & Andrus, D. N. (2022). An Exploration of Organizational Development and Change in Technical Organizations to Support Sustainability. International Journal of Smart Education and Urban Society (IJSEUS), 13(1), 1–14.

Burrell, D. N., Nobles, C., Cusak, A., Omar, M., & Gillesania, L. (2022). Cybercrime and the Nature of Insider Threat Complexities in Healthcare and Biotechnology Engineering Organizations. Journal of Crime and Criminal Behavior, 2(2), 131–144.

Wright, J., Dawson, M. E., Jr, & Omar, M. (2012). Cyber security and mobile threats: The need for antivirus applications for smart phones. *Journal of Information Systems Technology and Planning*, *5*(14), 40–60.

Omar, M., & Dawson, M. (2013). *Research in progress-defending android smartphones from malware attacks*. IEEE.

Dawson, Maurice, Al Saeed, I., Wright, J., & Omar, M. (2013). *Technology enhanced learning with open source software for scientists and engineers*. IATED.

Omar, M. (2012). *Smartphone Security: Defending Android-based Smartphone Against Emerging Malware Attacks*. Colorado Technical University.

Dawson, Maurice, Omar, M., & Abramson, J. (2015). Understanding the methods behind cyber terrorism. In *Encyclopedia of Information Science and Technology, Third Edition* (pp. 1539–1549). IGI Global.

Fawzi, D. R. A. J., & Omar, M. (n.d.). *NEW INSIGHTS TO DATABASE SECURITY AN EFFECTIVE AND INTEGRATED APPROACH TO APPLYING ACCESS CONTROL MECHANISMS AND CRYPTOGRAPHIC CONCEPTS IN MICROSOFT ACCESS ENVIRONMENTS*.

Dawson, Maurice, Omar, M., Abramson, J., & Bessette, D. (2014). The future of national and international security on the internet. In *Information security in diverse computing environments* (pp. 149–178). IGI Global.

Omar, M. (2015b). Insider threats: Detecting and controlling malicious insiders. In *New Threats and Countermeasures in Digital Crime and Cyber Terrorism* (pp. 162–172). IGI Global.

Dawson, Maurice, Wright, J., & Omar, M. (2015). Mobile devices: The case for cyber security hardened systems. In *New Threats and Countermeasures in Digital Crime and Cyber Terrorism* (pp. 8–29). IGI Global.

Dawson, Maurice. (2015). *New threats and countermeasures in digital crime and cyber terrorism*. IGI Global.

Hamza, Y. A., & Omar, M. D. (2013). Cloud computing security: abuse and nefarious use of cloud computing. *Int. J. Comput. Eng. Res*, *3*(6), 22–27.

Omar, M. (2015a). Cloud Computing Security: Abuse and Nefarious Use of Cloud Computing. In *Handbook of Research on Security Considerations in Cloud Computing* (pp. 30–38). IGI Global.

Davis, L., Dawson, M., & Omar, M. (2016). Systems Engineering Concepts with Aid of Virtual Worlds and Open Source Software: Using Technology to Develop Learning Objects and Simulation Environments. In *Handbook of Research on 3-D Virtual Environments and Hypermedia for Ubiquitous Learning* (pp. 483–509). IGI Global.

Mohammed, D., Omar, M., & Nguyen, V. (2017). Enhancing Cyber Security for Financial Industry through Compliance and Regulatory Standards. In *Security Solutions for Hyperconnectivity and the Internet of Things* (pp. 113–129). IGI Global.

Dawson, Maurice, Omar, M., Abramson, J., Leonard, B., & Bessette, D. (2017). Battlefield Cyberspace: Exploitation of Hyperconnectivity and Internet of Things. In *Developing Next-Generation Countermeasures for Homeland Security Threat Prevention* (pp. 204–235). IGI Global.

Nguyen, V., Omar, M., & Mohammed, D. (2017). A Security Framework for Enhancing User Experience. *International Journal of Hyperconnectivity and the Internet of Things (IJHIoT)*, *1*(1), 19–28.

Omar, M., Mohammed, D., & Nguyen, V. (2017). Defending against malicious insiders: a conceptual framework for predicting, detecting, and deterring malicious insiders. *International Journal of Business Process Integration and Management*, *8*(2), 114–119.

Dawson, Maurice, Eltayeb, M., & Omar, M. (2016). *Security solutions for hyperconnectivity and the Internet of things*. IGI Global.

Omar, M. (n.d.-b). *Latina Davis Morgan State University 1700 E Cold Spring Ln. Baltimore, MD 21251, USA E-mail: latinaedavis@ hotmail. com*.

Dawson, Maurice, Davis, L., & Omar, M. (2019). Developing learning objects for engineering and science fields: using technology to test system usability and interface design. *International Journal of Smart Technology and Learning*, *1*(2), 140–161.

Banisakher, M., Mohammed, D., & Omar, M. (2018). A Cloud-Based Computing Architecture Model of Post-Disaster Management System. *International Journal of Simulation--Systems, Science & Technology*, *19*(5).

Nguyen, V., Mohammed, D., Omar, M., & Banisakher, M. (2018). The Effects of the FCC Net Neutrality Repeal on Security and Privacy. *International Journal of Hyperconnectivity and the Internet of Things (IJHIoT)*, *2*(2), 21–29.

Omar, M., Mohammed, D., Nguyen, V., Dawson, M., & Banisakher, M. (2021). Android application security. In *Research Anthology on Securing Mobile Technologies and Applications* (pp. 610–625). IGI Global.

Banisakher, M., Omar, M., & Clare, W. (2019). Critical Infrastructure-Perspectives on the Role of Government in Cybersecurity. *Journal of Computer Sciences and Applications*, *7*(1), 37–42.

Omar, M. (2019). *A world of cyber attacks (a survey)*.

Mohammed, D., Omar, M., & Nguyen, V. (2018). Wireless sensor network security: approaches to detecting and avoiding wormhole attacks. *Journal of Research in Business, Economics and Management*, *10*(2), 1860–1864.

Banisakher, M., Omar, M., Hong, S., & Adams, J. (2020). A human centric approach to data fusion in post-disaster management. *J Business Manage Sci*, *8*(1), 12–20.

Nguyen, V., Mohammed, D., Omar, M., & Dean, P. (2020). *Net neutrality around the globe: A survey*. IEEE.

Dawson, M., Omar, M., Abramson, J., & Bessette, D. (2014). *INFORMATION SECURITY IN DIVERSE COMPUTING ENVIRONMENTS*.

Zangana, H. M., & Omar, M. (2020). Threats, Attacks, and Mitigations of Smartphone Security. *Academic Journal of Nawroz University*, *9*(4), 324–332.

Omar, M. (2021b). *New insights into database security: An effective and integrated approach for applying access control mechanisms and cryptographic concepts in Microsoft Access environments*.

Omar, M. (2021a). *Developing Cybersecurity Education Capabilities at Iraqi Universities*.

Omar, M., Gouveia, L. B., Al-Karaki, J., & Mohammed, D. (2022). Reverse-Engineering Malware. In *Cybersecurity Capabilities in Developing Nations and Its Impact on Global Security* (pp. 194–217). IGI Global.

Omar, M., Choi, S., Nyang, D., & Mohaisen, D. (2022b). Robust natural language processing: Recent advances, challenges, and future directions. *IEEE Access*.

Al Kinoon, M., Omar, M., Mohaisen, M., & Mohaisen, D. (2021). *Security breaches in the healthcare domain: a spatiotemporal analysis*. Springer International Publishing.

Omar, M. (2022b). *Machine Learning for Cybersecurity: Innovative Deep Learning Solutions. Springer Brief*. https://link.springer.com/book/978303115.

Omar, M. (n.d.-a). *Defending Cyber Systems through Reverse Engineering of Criminal Malware Springer Brief*. https://link.springer.com/book/9783031116278.

Omar, M., Choi, S., Nyang, D., & Mohaisen, D. (2022a). *Quantifying the Performance of Adversarial Training on Language Models with Distribution Shifts*.

Omar, M., & Mohaisen, D. (2022). *Making Adversarially-Trained Language Models Forget with Model Retraining: A Case Study on Hate Speech Detection*.

Omar, M. (2022c). Malware Anomaly Detection Using Local Outlier Factor Technique. In *Machine Learning for Cybersecurity: Innovative Deep Learning Solutions* (pp. 37–48). Springer International Publishing Cham.

Omar, M. (2022a). Application of Machine Learning (ML) to Address Cybersecurity Threats. In *Machine Learning for Cybersecurity: Innovative Deep Learning Solutions* (pp. 1–11). Springer International Publishing Cham.

Burrell, D. N., Nobles, C., Cusak, A., Omar, M., & Gillesania, L. (2022). Cybercrime and the Nature of Insider Threat Complexities in Healthcare and Biotechnology Engineering Organizations. *Journal of Crime and Criminal Behavior*, *2*(2), 131–144.

Omar, M., & Sukthankar, G. (2023). *Text-Defend: Detecting Adversarial Examples using Local Outlier Factor*. IEEE.

Ahmed, A., Rasheed, H., Bashir, A. K., & Omar, M. (2023). Millimeter-wave channel modeling in a VANETs using coding techniques. *PeerJ Computer Science*, 9, e1374.

Omar, M. (2023). *VulDefend: A Novel Technique based on Pattern-exploiting Training for Detecting Software Vulnerabilities Using Language Models*. IEEE.

Abbasi, R., Bashir, A. K., Mateen, A., Amin, F., Ge, Y., & Omar, M. (2023). Efficient Security and Privacy of Lossless Secure Communication for Sensor-based Urban Cities. *IEEE Sensors Journal*.

Shiaeles, M. O. A. (2023). *VulDetect: A novel technique for detecting software vulnerabilities using Language Models*. https://ieeexplore.ieee.org/document/10224924.

Gholami, S., & Omar, M. (2023c). Do Generative Large Language Models need billions of parameters? *arXiv Preprint arXiv:2309.06589*.

Burrell, D. N., Nobles, C., Richardson, K., Wright, J. B., Jones, A. J., Springs, D., … Brown-Jackson, K. (2023). Allison Huff. *Applied Research Approaches to Technology, Healthcare, and Business*, 1.

Huff, A. J., Burrell, D. N., Nobles, C., Richardson, K., Wright, J. B., Burton, S. L., … Brown-Jackson, K. L. (2023). Management Practices for Mitigating Cybersecurity Threats to Biotechnology Companies, Laboratories, and Healthcare Research Organizations. In *Applied Research Approaches to Technology, Healthcare, and Business* (pp. 1–12). IGI Global.

Gholami, S., & Omar, M. (2023a). Can a student Large Language Model perform as well as it's teacher? *arXiv Preprint arXiv:2310.02421*.

Ayub, M. F., Li, X., Mahmood, K., Shamshad, S., Saleem, M. A., & Omar, M. (2023). Secure Consumer-Centric Demand Response Management in Resilient Smart Grid as Industry 5.0 Application With Blockchain-Based Authentication. *IEEE Transactions on Consumer Electronics*.

Gholami, S., & Omar, M. (2023b). Can pruning make Large Language Models more efficient? *arXiv Preprint arXiv:2310.04573*.

Zhou, S., Ali, A., Al-Fuqaha, A., Omar, M., & Feng, L. (n.d.). Robust Risk-Sensitive Task Offloading for Edge-Enabled Industrial Internet of Things. *IEEE Transactions on Consumer Electronics*.

Gholami, S., & Omar, M. (2023d). Does Synthetic Data Make Large Language Models More Efficient? *arXiv Preprint arXiv:2310.07830*.

Tiwari, N., Omar, M., & Ghadi, Y. (2023). Brain Tumor Classification From Magnetic Resonance Imaging Using Deep Learning and Novel Data

Augmentation. In *Transformational Interventions for Business, Technology, and Healthcare* (pp. 392–413). IGI Global.

Tiwari, N., Ghadi, Y., & Omar, M. (2023). Analysis of Ultrasound Images in Kidney Failure Diagnosis Using Deep Learning. In *Transformational Interventions for Business, Technology, and Healthcare* (pp. 45–74). IGI Global.

Omar, M., Jones, R., Burrell, D. N., Dawson, M., Nobles, C., Mohammed, D., & Bashir, A. K. (2023). Harnessing the Power and Simplicity of Decision Trees to Detect IoT Malware. In *Transformational Interventions for Business, Technology, and Healthcare* (pp. 215–229). IGI Global.

Saleem, M. A., Li, X., Mahmood, K., Shamshad, S., Ayub, M. F., Bashir, A. K., & Omar, M. (2023). Provably Secure Conditional-Privacy Access Control Protocol for Intelligent Customers-centric Communication in VANET. *IEEE Transactions on Consumer Electronics*.

Omar, M., & Burrell, D. (2023). From text to threats: A language model approach to software vulnerability detection. *International Journal of Mathematics and Computer in Engineering*.

Arulappan, A., Raja, G., Bashir, A. K., Mahanti, A., & Omar, M. (2023). ZTMP: Zero Touch Management Provisioning Algorithm for the On-boarding of Cloud-native Virtual Network Functions. *Mobile Networks and Applications*, 1–13.

Al-Karaki, J. N., Omar, M., Gawanmeh, A., & Jones, A. (2023). *Advancing CyberSecurity Education and Training: Practical Case Study of Running Capture the Flag (CTF) on the Metaverse vs. Physical Settings*. IEEE.

Al Harthi, M. A. S., Al Balushi, M. M. Y., Al Badi, M. A. H., Al Karaki, J., & Omar, M. (n.d.). *Metaverse Adoption in UAE Higher Education: A Hybrid SEM-ANN Approach.......... 98 Mohammad Daradkeh, Boshra Aldhanhani, Amjad Gawanmeh, Shadi Atalla and Sami Miniaoui*.

Umer, M., Aljrees, T., Karamti, H., Ishaq, A., Alsubai, S., Omar, M., … Ashraf, I. (2023). Heart failure patients monitoring using IoT-based remote monitoring system. *Scientific Reports*, *13*(1), 19213.